\documentclass[manuscript]{aastex63}

\usepackage{txfonts}
\usepackage{latexsym,bm}
\newcommand{\qinemail}{qingang@hit.edu.cn}
\newcommand{\hit}{School of Science, Harbin Institute of
Technology, Shenzhen, 518055, China}
\newcommand{\hitqin}{\hit; \qinemail}
\usepackage{multirow}

\submitjournal{ApJ}

\shortauthors{Wu and Qin}

\watermark{Accepted}

\begin{document}

\title{Magnetic Cloud and Sheath in the Ground-Level Enhancement Event of 
2000 July 14. I. Effects on the Solar Energetic Particles}

\correspondingauthor{G. Qin}
\email{\qinemail}

\author[0000-0002-5776-455X]{S.-S. Wu}
\affiliation{\hitqin}

\author[0000-0002-3437-3716]{G. Qin}
\altaffiliation{Author of correspondence.}
\affiliation{\hitqin}

\begin{abstract}
Ground-level enhancements (GLEs) generally accompany with fast 
interplanetary coronal mass ejections (ICMEs), the shocks driven by which are 
the effective source of solar energetic particles (SEPs). In the GLE event of
2000 July 14, observations show that a very fast and strong magnetic cloud (MC)
is behind the ICME shock and the proton intensity-time profiles observed at 1 au
had a rapid two-step decrease near the sheath and MC. Therefore, we study the
effect of sheath and MC on SEPs accelerated by an ICME shock through numerically
solving the focused transport equation. The shock is regarded as a moving source
of SEPs with an assumed particle distribution function. The sheath and MC are
set to thick spherical caps with enhanced magnetic field, and the turbulence
levels in sheath and MC are set to be higher and lower than that of the ambient
solar wind, respectively. The simulation results of proton intensity-time
profiles agree well with the observations in energies ranging from $\sim$1 to
$\sim$100 MeV, and the two-step decrease is reproduced when the sheath and MC
arrived at the Earth. The simulation results show that the sheath-MC structure
reduced the proton intensities for about 2 days after shock passing through the
Earth. It is found that the sheath contributed most of the decrease while the
MC facilitated the formation of the second step decrease. The simulation also
infers that the coordination of magnetic field and turbulence in sheath-MC
structure can produce a stronger effect of reducing SEP intensities.
\end{abstract}


\section{Introduction}

The solar energetic particle (SEP) events, especially ground-level enhancements 
(GLEs), are one of the sources of space radiation harmful to the 
safety of spacecraft and the health of astronauts \citep[e.g.,][]{Lanzerotti17,
MertensEA18, MertensEA19}. Therefore, it is important to study the acceleration
and propagation of SEPs both in observation and theory. Over the
past several decades research in this field has made significant progress.

From the observation characteristics SEP events can be divided into two
categories: impulsive and gradual ones \citep[e.g.,][]{CaneEA86, Reames99, 
Reames17, Cliver09}. Impulsive events are believed to be caused by solar flares
with low intensity and short duration. On the other hand, gradual events are 
related to the shocks which are driven by coronal mass ejections (CMEs) and can
continuously release particles from corona to interplanetary space, so that they
usually last longer with higher flux. In addition, each of the two categories
can be further divided into two sub-categories according to recent research
\citep{Reames20}.

Based on the classification according to observations, numerical simulations for
SEP events with either impulsive sources or continuous sources are carried out
\citep[e.g.,][]{Droege00, ZhangEA09, DresingEA12, QinAWang15, QiEA17, HuEA18}.
The modeling work, focusing on the transport of SEPs in interplanetary space,
can be used to deal with a lot of problems such as the effects of adiabatic
cooling \citep[e.g.,][]{QinEA06}, perpendicular diffusion
\citep[e.g.,][]{ZhangEA09, WangEA12, DresingEA12}, the reservoir phenomenon
\citep[e.g.,][]{ZhangEA09, QinEA13}, and the release time of SEPs near the Sun
\citep[e.g.,][]{DiazEA11, WangAQin15}. 

Most of GLE events belong to the gradual category, usually accompanying with
fast interplanetary coronal mass ejections (ICMEs) \citep{GopalswamyEA12} that
are the interplanetary counterpart of CMEs \citep{LuhmannEA20}. ICMEs are 
macro-scale structures, thereby being able to drive many types of space weather
disturbances, such as geomagnetic storms and Forbush decreases (Fds) in galactic
cosmic ray (GCR) intensities \citep{Cane00, RichardsonEA11, Gopalswamy16}.
The ICMEs can be identified by specific plasma, compositional, and magnetic field
signatures. Particularly, if the signatures exhibit strong and smooth magnetic
field, coherent rotation of the magnetic field components, and low proton
temperature and plasma $\beta$ values, one can identify a magnetic cloud (MC)
embedded in the ICME \citep{BurlagaEA81, vanEA09, RichardsonEA10}. ICMEs can drive
shocks if their speed is sufficiently faster than the preceding solar wind, and
the shock is an effective accelerator of charged particles thus producing the
gradual SEP event. The region between the shock and ICME's leading edge is called
sheath, in which the turbulence level is greater than that in the ambient solar
wind due to the fact that the magnetic field lines are highly compressed by the
ICME and shock.

Many research have demonstrated that MCs and turbulent sheath regions can 
cause Fds \citep[e.g.,][]{ZhangEA88, Cane93, YuEA10, JordanEA11, RichardsonEA11,
LuoEA17,LuoEA18} as the result of modulating the intensity of GCRs, so that the 
other type of energetic particles, SEPs, accelerated by ICME-driven shocks may
also be significantly affected by MCs and sheath regions. Therefore, to evaluate
the influence of MC and sheath one can better predict SEP intensities.
Consequently, based on the prediction of solar activity
\citep[e.g.,][]{Petrovay10, QinAWu18}, the prediction of the trend of GLE events
on solar cycle scale, which is important for preventing major radiation hazards,
can be promoted \citep[e.g.,][]{Miroshnichenko18, WuAQin18}.

In this paper, we use the numerical code denoted as Shock Particle Transport 
Code (SPTC) developed by \citet{WangEA12} based on a stochastic differential
equation approach \citep{Zhang99, QinEA06} to study the effects of MC and sheath 
on SEPs released by an ICME shock, the simulation results will be compared with
the observations of GLE59, which occurred on 2000 July 14 and was accompanied 
with a very fast and strong MC \citep{Lepping01}. In Section~\ref{sec:obs}, the
observational features of GLE59 are presented, while the simulation model is
elaborated in Section~\ref{sec:model}. We show our simulation results and compare
them with observations in Section~\ref{sec:result}. Conclusions and discussion
are presented in Section~\ref{sec:discussion}. Note that, besides this work,
we use the same MC and Sheath model to study the Fd occurred following the GLE59,
and  reproduce the observed Fd successfully \citep{QinAWu20}.


\section{OBSERVATIONS}
\label{sec:obs}

\begin{figure}
\epsscale{1} \plotone{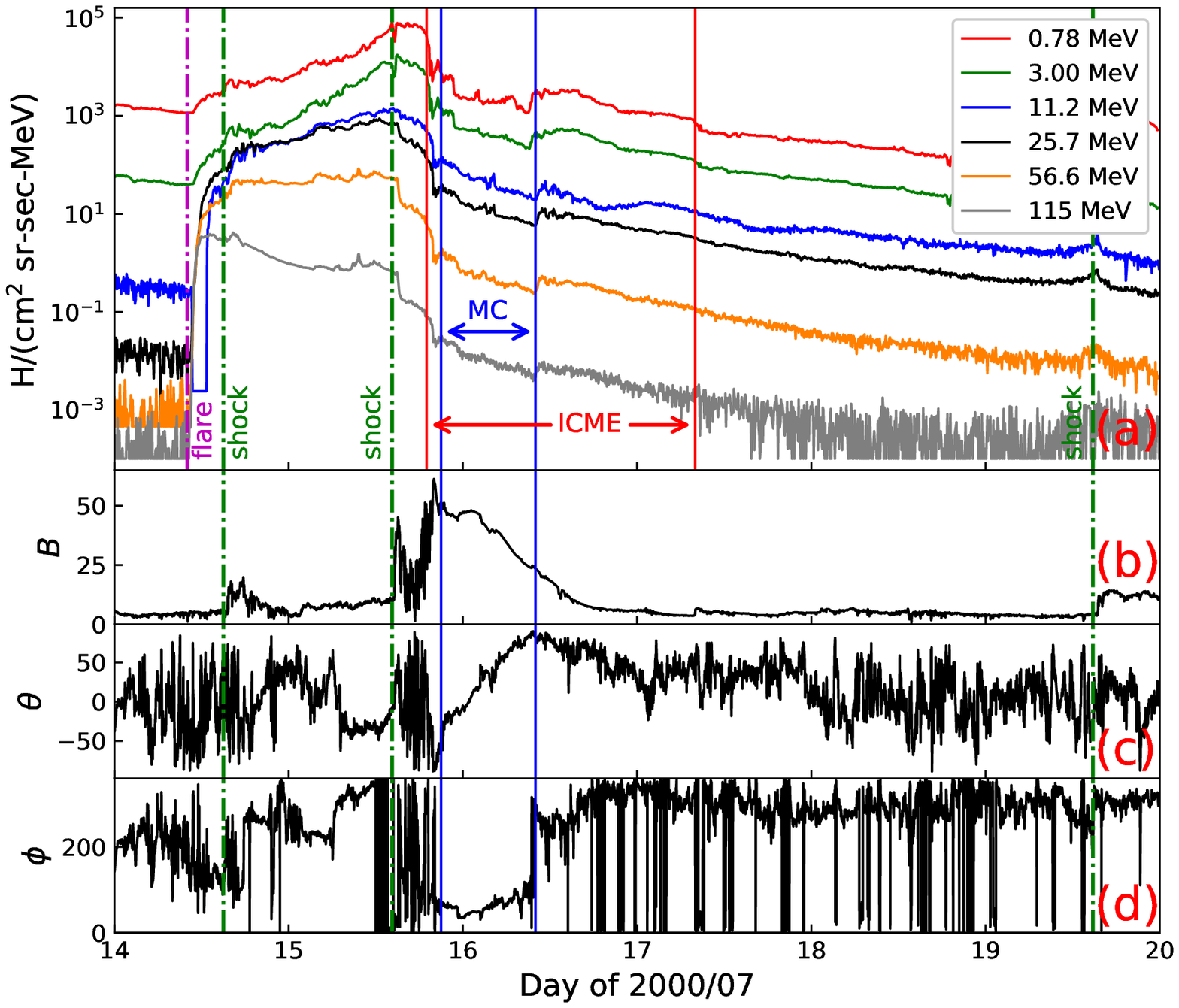}
\caption{Observations for GLE59. (a) The proton intensity-time profiles are 
observed near the Earth for six energy channels ranging from $\sim$1 to $\sim$100
MeV. The pink and green vertical dashed lines denote the flare onset and the
passages of ICME shocks, respectively. The boundaries of ICME and MC are presented
with the red and blue vertical solid lines, respectively. (b)$-$(d) are the
intensity, polar angle, and azimuthal angle of IMF in GSE angular coordinates,
respectively.}
\label{fig:observation}
\end{figure}

The proton intensity-time profiles of GLE59, the fifth GLE event in solar cycle 23,
are exhibited in Figure~\ref{fig:observation}(a) with observations from the
Electron, Proton, and Alpha Monitor (EPAM) \citep{GoldEA98} onboard \emph{ACE} and
Energetic Particle Sensor (EPS) \citep{OnsagerEA96} onboard \emph{GOES 8}.
Figures~\ref{fig:observation}(b)$-$\ref{fig:observation}(d) present the intensity,
polar angle, and azimuthal angle of interplanetary magnetic field (IMF) in GSE
angular coordinates observed by the Magnetic Field Investigation (MFI)
\citep{Lepping95} onboard \emph{Wind} spacecraft. In
Figure~\ref{fig:observation}(a), there was an X$5.7$ class flare that began at
$10$:$10$ UT on 14 July 2000 indicated by a pink vertical dashed line, and the
flare located at N$22$W$07$. There are three green vertical dashed 
lines denoting the passages of interplanetary shocks, and the second shock that 
arrived at $14$:$15$ UT on 15 July corresponds to the solar eruption. The 
arrival and departure times of the ICME were $19$:$00$ UT on 15 July and 
$8$:$00$ UT on 17 July, respectively, which are indicated by the two red 
vertical solid lines. The two blue vertical solid lines show the boundaries 
of the MC at $21$:$00$ UT 15 July and $10$:$00$ UT 16 July, respectively. 
When the second shock arrived at 1 au, the proton intensities showed a
significant enhancement, which is called the energetic storm particle event 
\citep[e.g.,][]{RaoEA68}. Subsequently, the proton intensities declined rapidly
until the MC passed through the Earth, after which the proton intensities 
recovered a little and decayed slowly finally.

Figure~\ref{fig:observation} exhibits that the proton intensities had a rapid 
decline phase after the shock arriving at the Earth and recovered a little
when the MC passed through the Earth later. The decline phase is similar to the
dropout phenomenon where particle intensity drops for a few hours and it is
usually observed in energies ranging from $\sim$0.02 to $\sim$5 MeV/nucleon for
ions \citep[e.g.,][]{MazurEA00, WangEA14, Tan17}. In GLE59, the decline phase can
be observed even in more than a hundred MeV protons, so that it is different from 
the dropout phenomenon. It is shown that the decline phase has a two-step 
decrease structure which a classical two-step Fd has \citep[e.g.,][]{Cane00} for 
high energy channels between the MC's leading edge and the shock, such as the 
gray and orange curves. The first step occurred right after the shock arrival,
which indicates that the first step might be caused by the turbulent sheath
region. Considering the fact that the second step appeared near the arrival time
of the MC and the recovery of proton intensities was close to the departure time
of the MC, we assume that the second step was caused by the MC. In the following,
we will reproduce the observed proton intensity-time profiles in simulation
considering the effects of sheath and MC that are placed behind an ICME shock.

The flare information is from \citet{GopalswamyEA12}, and the shock information
is from http://www-ssg.sr.unh.edu/mag/ace/ACElists/obs$\_$list.html$\#$shocks.
The start and end times of the ICME and MC are obtained from
\citet{RichardsonEA10}.

\section{SEP Transport MODEL}
\label{sec:model}
This section focuses on the simulation model, including the configurations of 
IMF, shock, MC, and sheath, transport equation, and diffusion coefficients.

\subsection{IMF, Shock, MC, and Sheath}

The Parker field is adopted as solar wind magnetic field and given by
\begin{equation}
\bm{B_{\text{P}}} = A B_{\text{P}0} \left( \frac{r_{\text{au}}}{r} \right)^2 
 \left(\bm{e_r} - \frac{\omega r \sin{\theta}}{V^{\text{sw}}} \bm{e_{\phi}}\right),
 \label{eq:BP}
\end{equation}
where $A=\pm1$ is the polarity, $B_{\text{P}0}$ is the radial component
of solar wind magnetic field at 1 au, $r_{\text{au}}$ is a constant and equals
to 1 au, $r$, $\theta$, and $\phi$ are the solar distance, polar angle, and
azimuthal angle of any point in a non-rotating heliographic coordinate system,
respectively, $\omega$ is the angular speed of solar rotation, and
$V^{\text{sw}}$ is solar wind speed.

In SPTC, the shock is treated as a spherical cap with uniform speed for 
releasing SEPs, and the longitude and latitude of shock nose are set to the 
same as those of the corresponding solar flare. The distribution function of 
the source $f_{\text{b}}(t, \bm{r})$ at time $t$ and position $\bm r$ is given
by the following equation
\citep{KallenrodeEA97, WangEA12}
\begin{equation}
f_{\text{b}} = f_0 \delta \left( r - v_{\text{s}} t \right) 
 \left( \frac{R_{\text{in}}}{r} \right)^{\alpha_p} 
 \exp{\left[ -\frac{\Omega \left( \theta,\phi \right)}{\Omega_p} \right] p^{\gamma_p}} 
 \qquad \left( \Omega \le \Omega_{\text{s}} \right),
 \label{eq:fb}
\end{equation}
where $f_0$ is a constant, $v_{\text{s}}$ is shock speed, $t$ is time,
$R_{in}$ is inner boundary, $\alpha_p$ and $\Omega_p$ are the
attenuation coefficients of shock strength in radial and angular directions,
$\Omega$ is the angular width from shock nose to the position $\bm r$,
$\Omega_{\text{s}}$ is the half angular width of shock, $p$ is the momentum
of particles, and $\gamma_p$ is spectral index that varies with $p$. The 
attenuation coefficients $\alpha_p$ and $\Omega_p$ are functions of the 
momentum of particles, i.e.,
\begin{eqnarray}
\alpha_p &=& \alpha_{0} \times \left( \frac{p}{p_{0}} \right)^{\eta_\alpha}, \\
\Omega_p &=& \Omega_{0} \times \left( \frac{p}{p_{0}} \right)^{\eta_\Omega},
\end{eqnarray}
where $\alpha_{0}$, $\Omega_{0}$, $\eta_\alpha$, $\eta_\Omega$, and $p_0$ are 
constants with $p_0=0.78$ MeV/$c$, and $c$ is the speed of light.
Note that, the spectral index $\gamma_p$ is not the energy spectrum
index since other parameters, i.e., $\alpha_p$ and $\Omega_p$ in
Equation~(\ref{eq:fb}) are also functions of energy.

\begin{figure}
\epsscale{1} \plotone{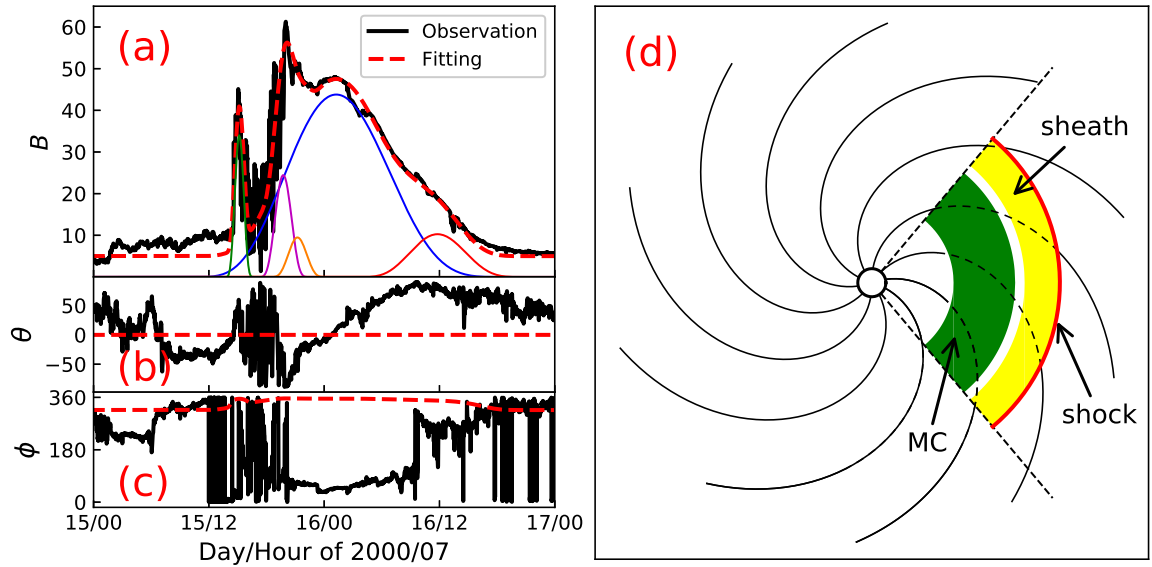}
\caption{(a) Fitting (red dashed line) of observed magnetic field (black solid
line) in sheath-MC structure with Equation~(\ref{eq:Bejecta}), and the red dashed
line is the sum of colored solid lines and Parker field at 1 au. The colored solid
lines are given by Equation~(\ref{eq:deltaB}), and the coefficients are listed in
Table~\ref{tab:para_fit}. 
(b)-(c) Comparisons between the polar and azimuthal angles of observed
IMF (black solid lines) and those of modeled IMF (red dashed lines), which are
calculated from the fitting result in Figure~\ref{fig:MC_model}a.}
(d) A sectional view of the IMF (black spiral curves), shock (red arc), sheath
(yellow area), and MC (green area) through the ecliptic plane.
\label{fig:MC_model}
\end{figure}

In this work, the MC and sheath are set as thick spherical caps behind the shock,
with the same direction, velocity, and angular width as those of the shock.
On the one hand,
Figure~\ref{fig:observation}(b) shows that the magnetic field in sheath-MC
structure is greater than that in the ambient solar wind, so that Parker field
is not suitable for representing the magnetic field in this area. 
On the other hand, it is hard for us to give a self-consistent
analytical magnetic field model to describe the complex three-dimensional
magnetic field in sheath-MC structure.
For simplicity,
the magnetic field in sheath-MC structure is set to the Parker field
$\bm B_{\text P}$ plus a magnetic field enhancement in radial
direction
\begin{equation}
\bm{B_{\text{ejecta}}} = \bm{B_{\text{P}}} + A \Delta B_r \bm{e_r},
 \label{eq:Bejecta}
\end{equation}
where $\Delta B_r$ can be expressed by the sum of a set of delta-like functions
and can be written as
\begin{eqnarray}
\Delta B_r &=& \sum_{i=1}^k \Delta B_{r}^i, \label{eq:deltaBr}\\
\Delta B_r^i &=& B_{r0}^i \left(\frac{r_{\text{au}}}{v_{\text{s}}t}\right)^2 
 \delta_n \left( \frac{v_{\text{s}} t + \delta r_i -r}{w_i} \right), \label{eq:deltaB}\\
\delta_n(x) &=& \left\{
\begin{array}{ll}
\left(1-x^2\right)^n & \qquad \text{for } x \in \left[-1,1\right],\\
0 & \qquad \text{for others},
\end{array} \right. \label{eq:deltan}
\end{eqnarray}
where $B_{r0}^i$, $\delta r_i$, $w_i$ ($i=1, 2, 3, ...,k$), and $n$ are constants  
obtained by fitting the observed magnetic field with Equation~(\ref{eq:Bejecta}).
Figure~\ref{fig:MC_model}(a) shows the fitting result, and the black solid and red
dashed curves are the observed and fitted magnetic field, respectively. The fitted
magnetic field is the sum of the magnetic enhancements represented by the colored
solid curves and Parker field. Each of the colored solid curves is given by
Equation~(\ref{eq:deltaB}), and the fitted coefficients are listed in
Table~\ref{tab:para_fit}. Note that, one can find the integral of the divergence
of magnetic field enhancement in sheath-MC structure along the radial direction
equals to zero, but $\nabla\cdot\left(\Delta B_r^i \bm{e_r}\right)$ is not zero in
most parts of the sheath-MC structure.
Panels(b)-(c) in Figure~\ref{fig:MC_model} present the comparison between
the observed and modeled polar and azimuthal angles of IMF with black solid and
red dashed lines. It is shown that the polar and azimuthal angles of the simplified
magnetic field in sheath-MC structure can not fit the observed ones due to the fact
that the observed magnetic field in sheath-MC is mostly in the azimuthal direction.
However, we use this simplified analytical magnetic field for the preliminary study
of the transport of SEPs focusing on the general characteristics of the effect of
the sheath-MC,
for example, the magnetic mirror effect
\citep[e.g.,][]{ReamesEA97, BieberEA02, TanEA09}.

Figure~\ref{fig:MC_model}(d) shows the sectional view of the IMF, shock, sheath, 
and MC through the ecliptic plane, represented by the black spiral curves, red
arc, thick yellow cap, and thick green cap, respectively. The spiral curves in
sheath-MC structure are plotted with dashed lines to indicate that the magnetic 
field is not Parker field in this area.

\subsection{Transport Equation}

We use the SPTC to model the transport of SEPs based on the previous studies
\citep[e.g.,][]{QinEA06, ZhangEA09, WangEA12}. The focused transport equation in
three-dimensional space is written as \citep{Skilling71, Schlicheiser02, QinEA06,
ZhangEA09}
\begin{eqnarray}
\frac{{\partial f}}{{\partial t}} + 
\left( v \mu \hat{\bm{b}} + \bm{V}^{\text{sw}} \right) \cdot \nabla f -
  \nabla \cdot \left( \bm{\kappa_\bot} \cdot \nabla f \right) - 
  \frac{\partial}{{\partial \mu}} 
  \left( D_{\mu \mu} \frac{{\partial f}}{{\partial \mu}} \right) \nonumber \\
- p \left[ {\frac{{1 - \mu^2}}{2} \left( {\nabla \cdot \bm{V}^{\text{sw}} -
  \hat{\bm{b}} \hat{\bm{b}} : \nabla \bm{V}^{\text{sw}}} \right) +
  \mu^2 \hat{\bm{b}} \hat{\bm{b}} : \nabla \bm{V}^{\text{sw}}} \right]
  \frac{{\partial f}}{{\partial p}} \nonumber \\
+ \frac{{1 - \mu^2}}{2} \left[ {-\frac{v}{L} + \mu \left( 
  \nabla \cdot \bm{V}^{\text{sw}} - 
  3 \hat{\bm{b}} \hat{\bm{b}} : \nabla \bm{V}^{\text{sw}} \right) } \right]
  \frac{{\partial f}}{{\partial \mu}}=0,
\label{eq:dfdt}
\end{eqnarray}
where $f(\bm{x},\mu,p,t)$ is the gyrophase-averaged distribution function and 
$\bm{x}$ is the particle position in a non-rotating heliographic coordinate
system, $v$ and $\mu$ are the speed and pitch-angle cosine of particles,
respectively, $\bm{V^{\text{sw}}} = V^{\text{sw}}\bm{e_r} $ is the solar wind
velocity, $\bm{\kappa_\bot}$ and $D_{\mu \mu}$ are the perpendicular and
pitch-angle diffusion coefficients of particles, respectively,
$L = \left( \hat{\bm{b}} \cdot \nabla \ln{B_0} \right)^{-1}$ is the magnetic
focusing length due to the in-homogeneous magnetic field, and $\hat{\bm{b}}$
is the unit vector along the local background magnetic field with strength $B_0$.
The equation includes the most of the particle transport mechanisms, i.e.,
particle streaming along magnetic field line and solar wind flowing in the IMF
(2nd term), perpendicular diffusion (3rd term), pitch-angle diffusion (4th term),
pitch-angle dependent adiabatic cooling by the expanding solar wind (5th term),
and focusing (6th term). We use a time-backward Markov stochastic process method
to solve Equation~(\ref{eq:dfdt}) \citep{Zhang99, QinEA06}.

\subsection{Diffusion Coefficient}
The model of pitch-angle diffusion coefficient $D_{\mu \mu}$ is set as 
\citep{BeeckEA86, TeufelEA03}
\begin{equation}
{D_{\mu \mu}}(\mu) = {\left( {\frac{{\delta {b_{\text{slab}}}}}{{{B_0}}}} \right)^2}
 \frac{{\pi (s-1)}}{{4s}} \frac{v}{l_{\text{slab}}} 
 \left(\frac{R_L}{l_{\text{slab}}}\right)^{s-2}
 \left({\mu^{s - 1}} + h \right) \left(1 - {\mu^2} \right),
 \label{eq:Dmumu}
\end{equation}
where $\delta b_{\text{slab}}$ is the slab component of magnetic turbulence,
$l_{\text{slab}}$ is the correlation length of $\delta b_{\text{slab}}$, $s=5/3$
is the Kolmogorov spectral index of the IMF turbulence in the inertial range, 
$R_L = pc / \left( |q| B_0 \right)$ is the Larmor radius with the charge of
particle $q$, and $h$ is a constant for modeling the non-linear effect of
pitch-angle diffusion at $\mu=0$. The parallel mean free path $\lambda_{||}$ is
given by
\citep{Jokipii66, HasselmannEA68, Earl74}
\begin{equation}
{\lambda_\parallel} = \frac{{3v}}{8} \int_{-1}^{+1} 
  {\frac{\left( 1 - {\mu^2} \right)^2}{D_{\mu \mu}} d \mu}.
\end{equation}

The perpendicular mean free path $\lambda_\bot$ is defined from the perpendicular
diffusion coefficient $\kappa_\perp$ for convenience
\begin{equation}
    \lambda_\perp\equiv \frac{3\kappa_\perp}{v},
\end{equation}
and from nonlinear guiding center theory
\citep{MatthaeusEA03} with analytical approximations \citep{ShalchiEA04, 
ShalchiEA10} we have
\begin{equation}
\lambda_\perp = 
{\left[ {{{\left( \frac{{\delta {b_{\text{2D}}}}}{{{B_0}}} \right)}^2}
\sqrt {3\pi } \frac{{s - 1}}{{2s}}
\frac{{\Gamma \left( \frac{s}{2} + 1 \right)}}
{{\Gamma \left( \frac{s}{2} + \frac{1}{2} \right)}}
{l_{2D}}} \right]^{2/3}} {\lambda_\parallel}^{1/3},
\label{eq:lambda_bot}
\end{equation}
where $\delta b_{\text{2D}}$ is the 2D component of magnetic turbulence, and 
$l_{\text{2D}}$ is the correlation length of $\delta b_{\text{2D}}$. In 
addition, the perpendicular diffusion coefficient $\bm{\kappa_{\perp}}$ can be
written as $\bm{\kappa_{\perp}} = \kappa_\perp \left( \bm{I} - \bm{\hat{b}}
\bm{\hat{b}} \right)$.

The turbulence level is given by
\begin{equation}
\sigma \equiv \frac{\delta b}{B_0} = 
 \frac{\sqrt{\delta b_{\text{slab}}^2 + \delta b_{\text{2D}}^2}}{B_{0}}.
\end{equation}
The ratio of 2D energy to slab energy is found to be 80\%:20\%
\citep{MatthaeusEA90, BieberEA94} and widely used in the literature
\citep[e.g.,][]{ZankEA92, ZankEA93, HunanaEA10}. By using the relation
$\delta b_{\text{2D}}^2 = 4 \delta b_{\text{slab}}^2$, the turbulence levels
of slab and 2D components in solar wind, sheath, and MC can be written as
\begin{eqnarray}
\left( \frac{\delta b_{\text{slab}}}{B_0} \right)_i &=& 
  \frac{\sqrt{5}}{5}\sigma_i  \qquad (i=\text{P, S, M}),\\
\left( \frac{\delta b_{\text{2D}}}{B_0} \right)_i &=& 
  \frac{2\sqrt{5}}{5}\sigma_i  \qquad (i=\text{P, S, M}),
\end{eqnarray}
where $\text{P, S, M}$ denote solar wind, sheath, and MC. Because the turbulence
level in MC and sheath is less and greater than that in solar wind, the value of
$\sigma_M$ and $\sigma_S$ should set to be lower and higher than that of
$\sigma_P$, respectively.

\section{SIMULATIONS AND COMPARISONS WITH OBSERVATIONS}
\label{sec:result}

\subsection{Parameter Settings}
\label{subsec:para}

Table~\ref{tab:para1} lists the main parameters in the simulations in this work. 
The half angular width of the shock, $\Omega_{\text{s}}$, is set to $45^\circ$.
The shock speed is set to 1406 km/s, which is calculated by dividing the Sun-Earth
distance by the shock transit time measured from the flare onset to the 1 au shock
arrival. In addition, 450 km/s is chosen as solar wind speed $V^{\text{sw}}$.
In order to make the solar wind magnetic field strength $B_{\text{P}}$ equal to
5 nT at 1 au, the radial strength of solar wind magnetic field at 1 au,
$B_{\text{P}0}$, is set to 3.62 nT. The half thickness of MC, $L_{\text{M}}$, is
set to 0.22 au, and the distance $d_{\text{M}}$ between the central position of
MC and the shock is set to 0.45 au, so that the arrival and departure times of MC
agree with the observations obtained by \citet{RichardsonEA10}. The half thickness
of sheath, $L_{\text{S}}$ is set to 0.08 au according to the passages of shock and
ICME's leading edge. Furthermore, the angular speed of solar rotation is set to
$\omega=2\pi/25.4$ rad/day, and the inner and outer boundaries of the simulation
are set to $R_{\text{in}}=0.05$ au and $R_{\text{out}}=50$ au, respectively.

The parameters of magnetic turbulence are listed in Table~\ref{tab:para2}. We set
$l_{\text{slab}}=0.025$ au, and thus $l_{\text{2D}}$ equals to
$l_{\text{slab}}/2.6=0.0096$ au according to the multi-spacecraft measurements
\citep[e.g.,][]{WeygandEA09, WeygandEA11}. The turbulence levels in solar wind,
sheath, and MC are set to 0.3, 1.6, and 0.1, respectively. The Kolmogorov spectral
index of the IMF turbulence, $s$, equals to 5/3 in the inertial range. And the
non-linear effect index, $h$, is set to 0.01.

The other shock parameters are obtained by fitting simulated proton 
intensity-time profiles to observations.
Firstly, the attenuation coefficient of shock strength in angular direction,
$\Omega_p$ has relatively low influence on proton time-intensity profiles, and thus
we give certain values manually for attenuation constant $\Omega_{0}$ and the
corresponding power-law index $\eta_\Omega$. Secondly, the attenuation coefficient
in radial direction, $\alpha_p$ is more important than other parameters to the shape
of proton time-intensity profiles, so that the attenuation constant $\alpha_0$ and
the corresponding power-law index $\eta_\alpha$ can be derived by fitting the shape
of simulated time-intensity profiles to that of the observed ones with a certain
spectral index $\gamma_p$. Thirdly, $\gamma_p$ is fitted for each energy channel
with the magnitude of simulated proton time-intensity profile and that of observed
one based on the derived $\alpha_0$ and $\eta_\alpha$. Finally, we fine-tune all
parameters to obtain the best fitting results.
The attenuation constants $\alpha_{0}$ and $\Omega_{0}$ are equal to 0.5 and
$10^\circ$, respectively, and the power-law indices $\eta_\alpha$ and $\eta_\Omega$
are equal to 0.86 and 0, respectively. The spectral index $\gamma_p$ for the six
energy channels as shown in Figure~\ref{fig:observation} equal to -10.0, -12.2,
-15.7, -20.6, -27.6, and -40.2, respectively.

\begin{figure}
\epsscale{0.6} \plotone{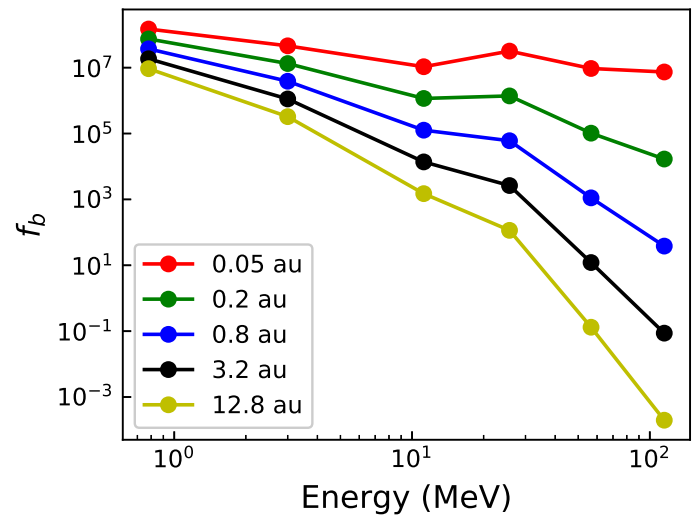}
\caption{The values of distribution function at shock nose are plotted
versus energy for different heliocentric distances.}
\label{fig:source}
\end{figure}

Figure~\ref{fig:source} shows the values of distribution function at
different heliocentric distances along the shock nose direction calculated by
using the fitted shock parameters. It is shown that the values of distribution
function can be represented by the power-law shape or the power-law shape with
an exponential tail, which is consistent with the result of shock acceleration
studies \citep[e.g.,][]{EllisonEA85, GiacaloneEA06, ZuoEA11, KongEA19}.

\subsection{Results}

\begin{figure}
\epsscale{1.1} \plotone{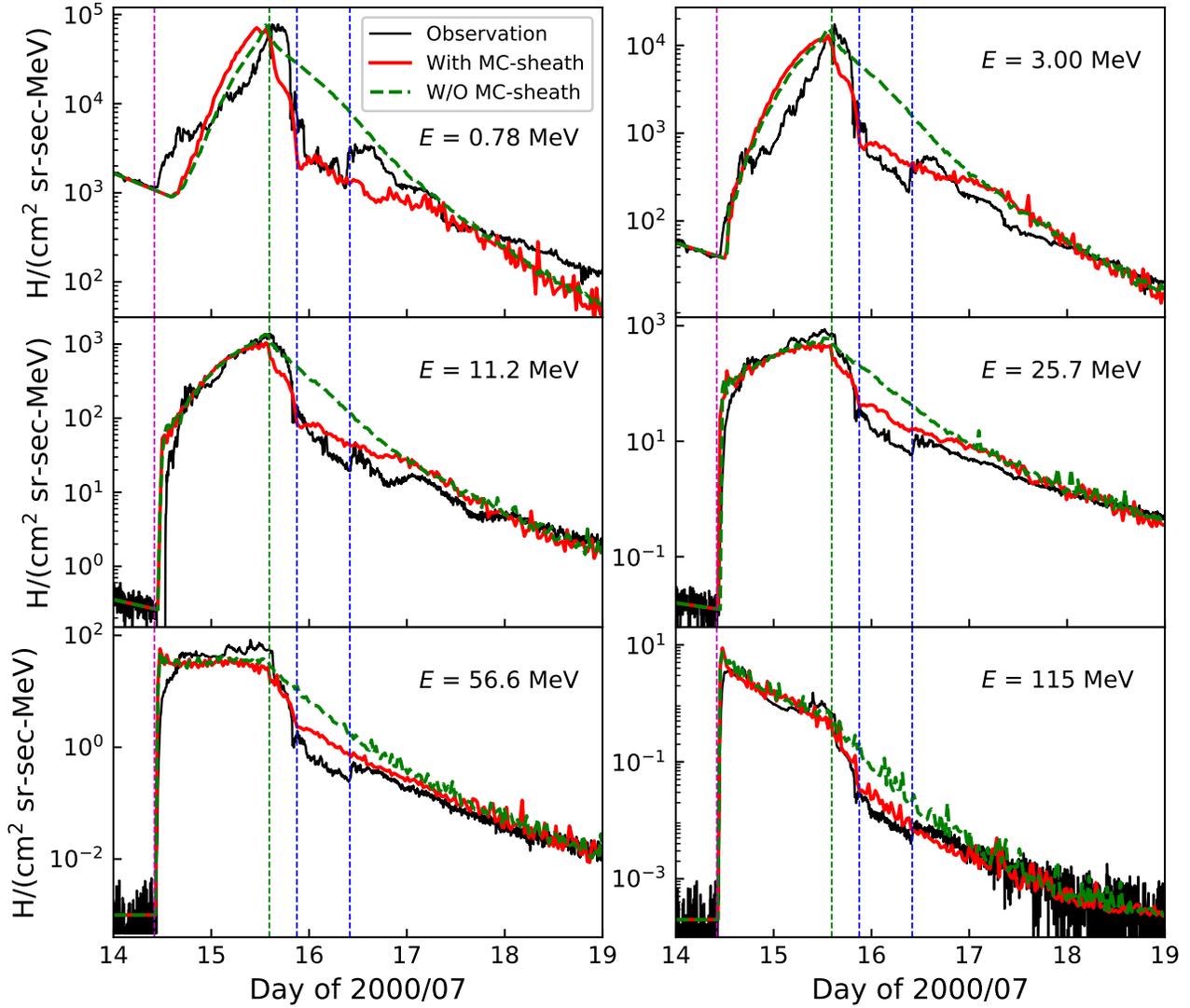}
\caption{Simulation results of GLE59 for six energy channels. The simulation
results with and without sheath-MC structure are presented by the red solid and
green dashed lines, respectively, and the black solid lines are the observations.
The pink, green, and blue vertical dashed lines denote the flare onset, shock
passage, and MC boundaries, respectively.}
\label{fig:simulation}
\end{figure}

The simulation results are presented in Figure~\ref{fig:simulation} where the
black solid curves are the observations while the red solid curves represent the
simulation results obtained by incorporating the MC and sheath into the SPTC. The
pink vertical line denotes the flare onset, before which the observed proton
intensities are used to obtain the backgrounds that are added to the simulated 
fluxes. The green vertical line represent the shock arrival while the blue vertical
lines show the arrival and departure times of MC. It is shown that the simulations
can fit the observations well, and the two-step decrease is reproduced between the
arrivals of shock and MC's leading edge.

To evaluate the effect of sheath-MC structure on SEPs, the simulation results
without MC and sheath are also presented in Figures~\ref{fig:simulation} by the
green dashed curves. It is shown that the shape of the green dashed curves is the
general one exhibited in the literature \citep[e.g.,][]{WangEA12, QinAQi20}
without rapid decline phase after the shock passage of the Earth. The
comparison between the green dashed and red solid curves shows the effect that the
sheath-MC structure can reduce SEP intensities when it arrived at the Earth, which
lasted about 2 days. The comparison also indicates that the sheath-MC structure
hardly affected the SEP intensities before it reached the Earth.

\begin{figure}
\epsscale{1.1} \plotone{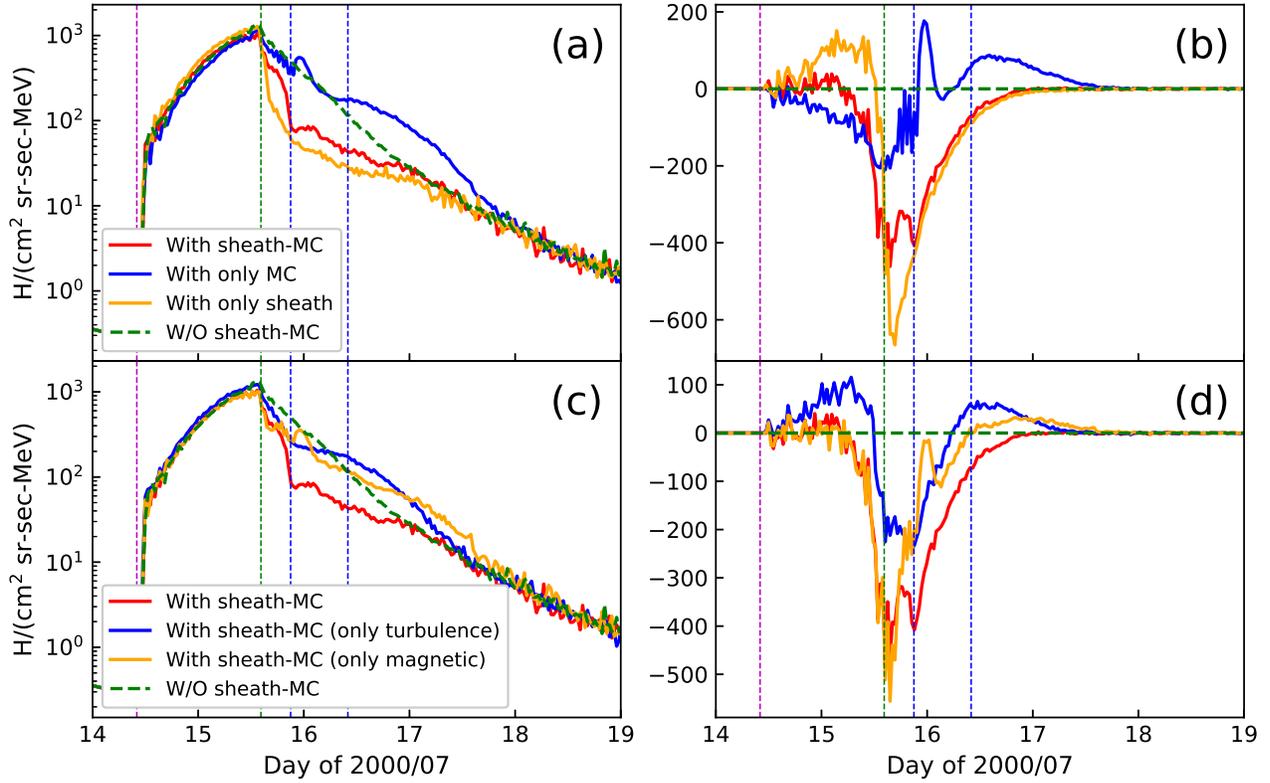}
\caption{Simulation results of GLE59 for 11.2 MeV protons. (a) The blue and
orange curves show the simulation results with only MC and sheath, respectively.
The other lines are the same as those in Figure~\ref{fig:simulation}. (b) The
differences of the four simulated lines and the green dashed line in
Figure~\ref{fig:simulation_compare}a. (c) The blue curve presents the simulation
result with sheath-MC structure where only turbulence level is different from
that in solar wind, while the orange curve shows the simulation result with
sheath-MC structure where only magnetic field is different from that in solar
wind. The other curves are the same as those in
Figure~\ref{fig:simulation_compare}a. (d) The same as
Figure~\ref{fig:simulation_compare}b but obtained from
Figure~\ref{fig:simulation_compare}c.}
\label{fig:simulation_compare}
\end{figure}

To further explore the respective influences of MC and sheath on SEPs, the
simulation results for 11.2 MeV protons with only MC or sheath are presented in
Figure~\ref{fig:simulation_compare}(a) by the blue or orange curves, respectively.
All the other lines in Figure~\ref{fig:simulation_compare}(a) are the same as
that in Figure~\ref{fig:simulation}. Figure~\ref{fig:simulation_compare}(b)
presents the difference between the four curves and the green curve in
Figure~\ref{fig:simulation_compare}(a). The simulation result with only MC, i.e.,
the blue curve is lower and higher than the green curve before and after the MC
arrival, respectively. It is also shown that, the simulation result with only
sheath, i.e., the orange curve is higher and lower than the green curve before
and after the shock arrival, respectively, which is similar to the
``diffusion barrier" effect \citep[e.g.,][]{LuoEA17, LuoEA18}. The simulation
result with sheath-MC, i.e., the red curve is almost always lower than the green
curve, and the first decrease is deeper than the second decrease. The
event-integrated fluences of the blue, orange, and red curves are 4\%, 23\%, and 
27\% less than that of the green curve, respectively. Therefore, the sheath plays
an important role in the decrease of SEP intensities, while the MC contributes to
the formation of the second step decrease.

It is also necessary to investigate the respective impacts of local background 
magnetic field and turbulence level on SEPs. Next, in our study we include the
sheath-MC structure. The simulation results with turbulence levels set according
to Section~\ref{subsec:para} with the local background magnetic fields the same as
that of ambient solar wind are plotted in Figure~\ref{fig:simulation_compare}(c)
by the blue curve. The simulation results with only the local background magnetic
fields different from that of ambient solar wind is presented by the orange curve
in Figure~\ref{fig:simulation_compare}(c). The other lines are the same as that in
Figure~\ref{fig:simulation_compare}(a). Figure~\ref{fig:simulation_compare}(d) has
the same format as Figure~\ref{fig:simulation_compare}(b) except that it is
originated from Figure~\ref{fig:simulation_compare}(c). It is clear that
neither the blue curve nor the orange curve can produce the two-step decrease. We
can show that the event-integrated fluences of the blue, orange, and red curves are
1\%, 16\%, and 27\% less than that of the green curve, respectively, which
indicates that the combination of turbulence and the enhancement of local
background magnetic field can produce stronger effect in the decrease of SEP
intensities.

\section{CONCLUSIONS AND DISCUSSION}
\label{sec:discussion}

In this work, we investigate the proton intensity-time profiles observed near the 
Earth for GLE59. It is shown that, the intensities have a rapid two-step decrease
after the ICME shock arriving at the Earth for $\sim$1 to $\sim$100 MeV protons,
which is clearer with higher energy. The two-step decrease is assumed to be caused
by the sheath region and MC. To reproduce the phenomenon, a simplified sheath-MC
structure is incorporated into the SPTC for simulating the transport of shock
accelerated energetic particles.

The shock is treated as a moving source of SEPs with uniform speed, which is
determined by dividing the Sun-Earth distance by the shock transit time, and
the longitude and latitude of shock nose are set to the same as that of the
corresponding solar flare. For simplicity, the MC and sheath are set as thick
spherical caps with the direction, speed, and angular width set as the
same as that of the shock. The Parker field is chosen as the local background
magnetic field in solar wind, while the Parker field plus a magnetic enhancement
represented by Equation~(\ref{eq:Bejecta}) is adopted as the local background
magnetic field in sheath-MC structure. Besides, the turbulence levels in MC and
sheath are set to lower and higher than that in solar wind, respectively.

The simulation results indicate that the observed proton intensity-time profiles
of GLE59 can be fitted well when a sheath-MC structure is placed behind the ICME 
shock for $\sim$1 to $\sim$100 MeV protons, and the two-step decrease of proton
intensities is reproduced when the sheath and MC's leading edge arrived at the 
Earth. Besides, the comparison between the simulation results with and without
sheath-MC structure shows that the sheath-MC structure hardly affected the proton
intensities before the shock arrived at the Earth, while it reduced the proton
intensities after the shock arrival. The reducing effect lasted about 2 days. The 
first decrease is found to be deeper than the second one. Furthermore,
The simulation results with only sheath or MC infers that the sheath contributed
most of the decrease and the MC played an important role in the formation of
the second step decrease. We also investigate that the effect of the combination
of the turbulence and local background magnetic field in sheath-MC structure is
greater than the simply superimposing of their respective effects on reducing SEP
intensities.

The simulation results show that the sheath contributed most of the decrease and
the first decrease is deeper than the second one, which can be explained by two
reasons. Firstly, the magnetic focusing effect occurs if magnetic field is
in-homogeneous, so that the strong focusing effects in sheath and MC due to the
rapid change of the local background magnetic fields acting as magnetic mirrors
to block the passage of particles, thus reducing SEP intensities. The magnetic
focusing effect in sheath is greater than that in MC because the magnetic field
in sheath varies faster than that in MC. Secondly, the turbulence level in sheath
is higher than that in MC, resulting in shorter parallel mean free path of
energetic particles in sheath than that in MC, and consequently the energetic
particle intensities will be reduced \citep[e.g.,][]{LuoEA17, LuoEA18}.

It is impossible to measure the time-varying magnetic field of overall space now.
For simplicity, the Parker field is chosen as the solar wind magnetic field and the
Parker field plus a magnetic field enhancement is used as the magnetic field in
sheath-MC structure. Furthermore, we use a simplified analytical magnetic field
enhancement in sheath-MC model, which is not divergence-free, for the preliminary
study of the transport of SEPs. 
The magnetic enhancement is set in radial direction for simplicity since
it is difficult for us to provide a self-consistent analytical three-dimensional
sheath-MC model, causing the modeled azimuthal and polar angles inconsistent with
observations. Since the energetic particles' diffusion process depends on the
direction of local magnetic field, the inaccuracy of magnetic enhancement direction
would make the simulation result different from observation. However, we think
the magnetic mirror effect from this simplified model is still successful to
reproduce the observational characteristics.
It is assumed that we can roughly describe the general characteristics of SEPs
flux affected by the sheath and MC with this model. In the future, we may adopt
an analytical sheath-MC structure without divergence instead. In addition, to
provide a local background magnetic field, we may use a three-dimensional
magnetohydrodynamic simulation \citep[e.g.,][]{LuoEA13, PomoellEA18, WijsenEA19}
with some divergence-free schemes \citep{BalsaraAKim04}.

In this work, the shock is treated as a moving source of energetic particles with
a pre-described distribution function Equation~(\ref{eq:fb}). According to the
fitted values of $\alpha_{0}$ and $\eta_{\alpha}$ in Table~\ref{tab:para1}, the
distribution function $f_{\text{b}}$ diminishes along the radial direction
gradually, and the decay is faster if the energy is higher. It is reasonable
because high energy particles are believed to be produced near the corona.
The research on the acceleration of energetic particles 
\citep[e.g.,][]{KongEA17, KongEA19, QinEA18, KongEA20} by shocks can provide a
more fidelity source, the including of which in this work may promote the
understanding of the effects of sheath and MC.

The parallel and perpendicular diffusion is described by well established
models, i.e., Equations~(\ref{eq:Dmumu}) $-$ (\ref{eq:lambda_bot}). However, the
turbulence parameters in these equations are simplified, for example, the radial
dependent is not considered. With the current progress in turbulence theory by the
community \citep[e.g.,][]{ZankEA18, ZhaoEA18, AdhikariEA20}, we can better
understand the transport of SEPs by using radial and time dependent turbulence
parameters. Besides, The \emph{Parker Solar Probe (PSP)} can provide near-Sun solar
wind and SEP observations \citep{BaleEA19, KasperEA19, McComasEA19} that can
promote the understanding of the radial evolution of turbulence, ICMEs, shocks,
and SEPs, which are important to our studies. Therefore, we will adopt the new
achievement about the understanding of solar wind turbulence in the future research.

\acknowledgments
This work was supported, in part, under grant NNSFC 41874206.
We thank the \emph{ACE} EPAM; \emph{GOES} EPS; \emph{Wind} MFI teams for
providing the data used in this paper. The \textit{ACE} data are provided by
the ACE Science Center and the \emph{GOES} data by the NOAA. We appreciate
the availability of the \emph{Wind} data at the Coordinated Data Analysis Web.
The work was carried out at National Supercomputer Center in Tianjin, and the
calculations were performed on TianHe-1 (A).




\clearpage

\clearpage

\clearpage

\clearpage

\clearpage

\clearpage
\begin{deluxetable}{ccccc}
\tablecaption{Parameters for representing magnetic field enhancements in
Figure~\ref{fig:MC_model}(a) with Equations~(\ref{eq:deltaBr}) $-$ (\ref{eq:deltan}).
\label{tab:para_fit}}
\tablehead{
\colhead{Color\tablenotemark{a}} & \colhead{$B_{r0}$} & \colhead{$\delta r$}
& \colhead{$w$}& \colhead{$n$}
}
\startdata
Green & 10 & 0.42 & 0.05 & 5\\
Pink & 10 & 0.26 & 0.09 & 5\\
Orange & 4 & 0.21 & 0.1 & 5\\
Blue & 24 & 0.03 & 0.56 & 5\\
Red & 9 & -0.29 & 0.3 & 5
\enddata
\tablenotetext{a}{It denotes the color of the colored solid curves in
Figure~\ref{fig:MC_model}(a).}
\end{deluxetable}

\clearpage
\begin{deluxetable}{cccc}
\tablecaption{Parameter settings for the simulation.
\label{tab:para1}}
\tablehead{
\colhead{Type} & \colhead{Parameter} & \colhead{Meaning} & \colhead{Value}
}
\startdata
\multirow{6}{*}{Shock} & $\Omega_{\text{s}}$ & half angular width & $45^{\circ}$\\
& $v_{\text{s}}$ & speed & 1406 km/s\\
& $\alpha_{0}$ & attenuation constant in radial & 0.5\\
& $\eta_\alpha$ & power-law index of $\alpha_p$ & 0.86\\
& $\Omega_{0}$ & attenuation constant in angular & 10$^\circ$\\
& $\eta_\Omega$ & power-law index of $\Omega_{p}$ & 0\\
\hline
\multirow{3}{*}{Solar wind} & $V^{\text{sw}}$ & speed & 450 km/s\\
& $B_{\text{P}0}$ & radial strength of IMF at 1 au & 3.62 nT\\
& $B_{\text{P}}|_{1\text{au}}$ & total strength of IMF at 1 au & 5 nT\\
\hline
\multirow{2}{*}{MC}	& $L_{\text{M}}$ & half thickness & 0.22 au\\
& $d_{\text{M}}$ & distance between MC center and shock & 0.45 au\\
\hline
\multirow{1}{*}{Sheath}	& $L_{\text{S}}$ & half thickness & 0.08 au\\
\hline
\multirow{3}{*}{Others} & $\omega$ & angular speed of solar rotation & 
  2$\pi$/25.4 \ rad/day\\
& $R_{\text{in}}$ & inner boundary of simulation & 0.05 au\\
& $R_{\text{out}}$ & outer boundary of simulation & 50 au\\
\enddata
\end{deluxetable}

\clearpage
\begin{deluxetable}{ccc}
\tablecaption{Parameter settings of turbulence for the simulation.
\label{tab:para2}}
\tablehead{
	\colhead{Parameter} & \colhead{Meaning} & \colhead{Value}
}
	\startdata
	$\sigma_{\text{P}}$ & turbulence level in solar wind & 0.3\\
	$\sigma_{\text{S}}$ & turbulence level in sheath & 1.6\\
	$\sigma_{\text{M}}$ & turbulence level in MC & 0.1\\
	$l_{\text{slab}}$ & slab correlation length & 0.025 au\\
	$l_{\text{2D}}$ & 2D correlation length & 0.0096 au\\
	$s$ & Kolmogorov spectral index &$5/3$\\
	$h$ &  non-linear effect index & $0.01$\\
	\enddata
\end{deluxetable}


\begin{thebibliography}{}

\bibitem[{Adhikari et~al.(2020)}]{AdhikariEA20}
Adhikari, L., Zank, G. P., Zhao, L.-L., et al. 2020, \apjs, 246, 38

\bibitem[{Bale et~al.(2019)}]{BaleEA19}
Bale, S. D., Badman, S. T., Bonnell, J. W., et al. 2019, \nat, 576, 237

\bibitem[{Balsara \& Kim(2004)}]{BalsaraAKim04}
{Balsara}, D. S., \& {Kim}, J. 2004, \apj, 602, 1079

\bibitem[{Beeck \& Wibberenz(1986)}]{BeeckEA86}
{Beek}, J., \& {Wibberenz}, G. 1986, \apj, 311, 437

\bibitem[{Bieber et~al.(2002)}]{BieberEA02}
Bieber, J. W., {Dr\"{o}ge}, W., Evenson, P. A., et al. 2002, \apj, 567, 622

\bibitem[{Bieber et~al.(1994)}]{BieberEA94}
Bieber, J. W., Matthaeus, W. H., Smith, C. W., et al. 1994, \apj, 420, 294

\bibitem[{Burlaga et~al.(1981)}]{BurlagaEA81}
Burlaga, L., Sittler, E., Mariani, F., et al. 1981, \jgr, 86, 6673

\bibitem[{{Cane}(1993)}]{Cane93}
{Cane}, H. V. 1993, \jgr, 98, 3509

\bibitem[{{Cane}(2000)}]{Cane00}
{Cane}, H. V. 2000, \ssr, 93, 55

\bibitem[{Cane et~al.(1986)}]{CaneEA86}
Cane, H. V., McGuire, R. E., \& von Rosenvinge, T. T. 1986, \apj, 301, 448

\bibitem[{{Cliver}(2009)}]{Cliver09}
{Cliver}, E. W. 2009, in Proc. IAU Symp. 257, Universal Heliophysical Processes,
ed. N. Gopalswamy \& D. F. Webb (Cambridge: Cambridge Univ. Press), 401

\bibitem[Diaz et~al.(2011)]{DiazEA11}
Diaz, I., Zhang, M., Qin, G., et al. 2011, ICRC, 10, 40

\bibitem[{{Dresing} {et~al.}(2012)}]{DresingEA12}
{Dresing}, N., {G{\'o}mez-Herrero}, R., {Klassen}, A., et al. 2012, \solphys, 281, 281

\bibitem[{{Dr\"{o}ge}(2000)}]{Droege00}
{Dr\"{o}ge}, W. 2000, \apj, 537, 1073

\bibitem[{{Earl}(1974)}]{Earl74}
{Earl}, J. A. 1974, \apj, 193, 231

\bibitem[{Ellison \& Ramaty(1985)}]{EllisonEA85}
{Ellison}, D. C., \& {Ramaty}, R. 1985, \apj, 298, 400

\bibitem[{Giacalone \& K{\'{o}}ta(2006)}]{GiacaloneEA06}
{Giacalone}, J., \& {K{\'{o}}ta}, J. 2006, \ssr, 124, 277

\bibitem[{Gold et~al.(1998)}]{GoldEA98}
Gold, R.~E., Krimigis, S.~M., Hawkins, S.~E.
  \uppercase\expandafter{\romannumeral3}, et al. 1998,
  \ssr, 86, 541

\bibitem[{{Gopalswamy}(2016)}]{Gopalswamy16}
{Gopalswamy}, N. 2016, GSL, 3, 8

\bibitem[{Gopalswamy et~al.(2012)}]{GopalswamyEA12}
Gopalswamy, N., Xie, H., Yashiro, S., et al. 2012, \ssr, 171, 23

\bibitem[{Hasselmann \& Wibberenz(1968)}]{HasselmannEA68}
{Hasselmann}, K., \& {Wibberenz}, G. 1968, ZGeo, 34, 353

\bibitem[{{Hu} {et~al.}(2018){Hu}, {Li}, {Fu}, {Zank}, \& {Ao}}]{HuEA18}
{Hu}, J., {Li}, G., {Fu}, S., et al. 2018, \apjl, 854, L19

\bibitem[{Hunana \& Zank(2010)}]{HunanaEA10}
{Hunana}, P., \& {Zank}, G. P. 2010, \apj, 718, 148

\bibitem[{{Jokipii}(1966)}]{Jokipii66}
{Jokippi}, J. R. 1966, \apj, 146, 480

\bibitem[{{Jordan et~al.}(2011)}]{JordanEA11}
Jordan, A. P., Spence, H. E., Blake, J. B., et al. 2011, 
\jgr, 116, A11103

\bibitem[{Kallenrode \& Wibberenz(1997)}]{KallenrodeEA97}
{Kallenrode}, M.-B., \& {Wibberenz}, G. 1997, \jgr, 102, A10, 22311

\bibitem[{{Kasper et~al.}(2019)}]{KasperEA19}
Kasper, J. C., Bale, S. D., Belcher, J. W., et al. 2019, 
\nat, 576, 228

\bibitem[{{Kong} \& {Qin} (2020)}]{KongEA20}
{Kong}, F.-J., \& {Qin}, G. 2020, \apj, 896, 20

\bibitem[{{Kong} {et~al.}(2019)}]{KongEA19}
{Kong}, F.-J., {Qin}, G., {Wu}, S.-S., et al. 2019, \apj, 877, 97

\bibitem[{{Kong} {et~al.}(2017){Kong}, {Qin}, \& {Zhang}}]{KongEA17}
{Kong}, F.-J., {Qin}, G., \& {Zhang}, L.-H. 2017, \apj, 845, 43

\bibitem[{{Lanzerotti}(2017)}]{Lanzerotti17}
{Lanzerotti}, L. J. 2017, \ssr, 212, 1253

\bibitem[{{Lepping} {et~al.}(1995)}]{Lepping95}
{Lepping}, R. P., Ac{\~{u}}na, M. H., {Burlaga}, L. F., et al. 1995, \ssr, 71, 207

\bibitem[{{Lepping} {et~al.}(2001)}]{Lepping01}
{Lepping}, R. P., Berdichevsky, D. B., \& {Burlaga}, L. F., et al. 2001, \solphys, 204, 287

\bibitem[{{Luhmann} {et~al.}(2020)}]{LuhmannEA20}
Luhmann, J. G., Gopalswamy, N., Jian, L. K., et al. 2020, \solphys, 295, 61

\bibitem[{{Luo} {et~al.}(2017)}]{LuoEA17}
{Luo}, X., {Potgieter}, M. S., {Zhang}, M., et al. 2017, \apj, 839, 53

\bibitem[{{Luo} {et~al.}(2018)}]{LuoEA18}
{Luo}, X., {Potgieter}, M. S., {Zhang}, M., et al. 2018, \apj, 860, 160

\bibitem[{{Luo} {et~al.}(2013)}]{LuoEA13}
Luo, X., Zhang, M., Rassoul, H. K., et al. 2013, \apj, 764, 85

\bibitem[{{McComas} {et~al.}(2019)}]{McComasEA19}
McComas, D. J., Christian, E. R., Cohen, C. M. S., et al. 2019, \nat, 576, 223

\bibitem[{{Matthaeus} {et~al.}(1990)}]{MatthaeusEA90}
{Matthaeus}, W. H., {Goldstein}, M. L., \& {Roberts}, D. A. 1990, \jgr, 95, 20673

\bibitem[{{Matthaeus} {et~al.}(2003){Matthaeus}, {Qin}, {Bieber}, 
\& {Zank}}]{MatthaeusEA03}
{Matthaeus}, W. H., {Qin}, G., {Bieber}, J. W., et al. 2003, 
\apj, 590, L53

\bibitem[{{Mazur} {et~al.}(2000)}]{MazurEA00}
{Mazur}, J. E., {Mason}, G. M., {Dwyer}, J. R., et al. 2000, \apj, 532, L79

\bibitem[{Mertens \& Slaba(2019)}]{MertensEA19}
{Mertens}, C. J., \& {Slaba}, T. C. 2019, SpWea, 17, 1650

\bibitem[{{Mertens} {et~al.}(2018)}]{MertensEA18}
Mertens, C. J., Slaba, T. C., \& Hu, S. 2018, SpWea, 16, 1291

\bibitem[{Miroshnichenko(2018)}]{Miroshnichenko18}
Miroshnichenko, L. I. 2018, JSWSC, 8, A52

\bibitem[{Onsager et~al.(1996)}]{OnsagerEA96}
Onsager, T. G., Grubb, R., Kunches, J., et al. 1996, \procspie, 2812, 281

\bibitem[{Petrovay(2010)}]{Petrovay10}
Petrovay, K. 2010, LRSP, 7, 6

\bibitem[{Pomoell \& Poedts(2018)}]{PomoellEA18}
{Pomoell}, J., \& {Poedts}, S. 2018, JSWSC, 8, A35

\bibitem[{Qi {et~al.}(2017)Qi, Qin, \& Wang}]{QiEA17}
Qi, S.-Y., Qin, G., \& Wang, Y. 2017, RAA, 17, 33

\bibitem[{{Qin} {et~al.}(2018)Qin, Kong, \& Zhang}]{QinEA18}
Qin, G., Kong, F.-J., \& Zhang, L.-H. 2018, \apj, 860, 3

\bibitem[{{Qin} \& {Qi}(2020)}]{QinAQi20}
{Qin}, G., \& {Qi}, S.-Y. 2020, \aap, 637, A48

\bibitem[{Qin \& Wang(2015)}]{QinAWang15}
{Qin}, G., \& {Wang}, Y. 2015, \apj, 809, 177

\bibitem[{{Qin} {et~al.}(2013){Qin}, {Wang}, {Zhang}, \&
  {Dalla}}]{QinEA13}
{Qin}, G., {Wang}, Y., {Zhang}, M., et al. 2013, \apj, 766, 74

\bibitem[{{Qin} \& {Wu}(2018)}]{QinAWu18}
{Qin}, G., \& {Wu}, S.-S. 2018, \apj, 869, 48

\bibitem[{{Qin} \& {Wu}(2020)}]{QinAWu20}
{Qin}, G., \& {Wu}, S.-S. 2020, \apj, to be submitted

\bibitem[{{Qin} {et~al.}(2006){Qin}, {Zhang}, \& {Dwyer}}]{QinEA06}
{Qin}, G., {Zhang}, M., \& {Dwyer}, J.~R. 2006, \jgr, 111, A08101

\bibitem[{Rao et~al.(1968)}]{RaoEA68}
Rao, U.~R., McCracken,  K.~G., \& Bukata, R.~P. 1968, CaJPh, 46, S844

\bibitem[{Reames et~al.(1997)}]{ReamesEA97}
Reames, D. V., Kahler, S. W., \& Ng, C. K. 1997, \apj, 491, 414

\bibitem[{Reames(1999)}]{Reames99}
Reames, D.~V. 1999, \ssr, 90, 413

\bibitem[{Reames(2017)}]{Reames17}
Reames, D.~V. 2017, Solar Energetic Particles (Berlin: Springer)

\bibitem[{Reames(2020)}]{Reames20}
Reames, D.~V. 2020, \ssr, 216, 20

\bibitem[{Richardson \& Cane(2010)}]{RichardsonEA10}
{Richardson}, I. G., \& {Cane}, H. V. 2010, \solphys, 264, 189

\bibitem[{Richardson \& Cane(2011)}]{RichardsonEA11}
{Richardson}, I. G., \& {Cane}, H. V. 2011, \solphys, 270, 609

\bibitem[{{Schlicheiser}(2002)}]{Schlicheiser02}
{Schlicheiser}, R. 2002, Cosmic Ray Astrophysics (Berlin: Springer)

\bibitem[{Shalchi et~al.(2004)}]{ShalchiEA04}
{Shalchi}, A., Bieber, J. W., Matthaeus, W. H., et al. 2004, \apj, 616, 617

\bibitem[{Shalchi et~al.(2010)}]{ShalchiEA10}
{Shalchi}, A., Li, G., \& {Zank}, G. P. 2010, \apss, 325, 99

\bibitem[{{Skilling}(1971)}]{Skilling71}
{Skilling}, J. 1971, \apj, 170, 265

\bibitem[{{Tan}(2017)}]{Tan17}
{Tan}, L. C. 2017, \apj, 846, 18

\bibitem[{Tan et~al.(2009)}]{TanEA09}
Tan, L. C., Reames, D. V., Ng, C. K., et al. 2009, \apj, 701, 1753

\bibitem[{Teufel \& Schlickeiser(2003)}]{TeufelEA03}
{Teufel}, A., \& {Schlickeiser}, R. 2003, \aap, 397, 15

\bibitem[{van Driel-Gesztelyi \& Culhane(2009)}]{vanEA09}
{van Driel-Gesztelyi}, L., \& {Culhane}, J. L. 2009, \ssr, 144, 351

\bibitem[{Wang \& Qin(2015)}]{WangAQin15}
{Wang}, Y., \& {Qin}, G. 2015, \apj, 806, 252

\bibitem[{{Wang} {et~al.}(2012){Wang}, {Qin}, \& {Zhang}}]{WangEA12}
{Wang}, Y., {Qin}, G., \& {Zhang}, M. 2012, \apj, 752, 37

\bibitem[{{Wang} {et~al.}(2014){Wang}, {Qin}, {Zhang}, \& {Dalla}}]{WangEA14}
{Wang}, Y., {Qin}, G., {Zhang}, M., et al. 2014, \apj, 789, 157

\bibitem[{{Weygand} {et~al.}(2009)}]{WeygandEA09}
{Weygand}, J. M., {Matthaeus}, W. H., {Dasso}, S., et al. 2009, \jgr, 114, A07213

\bibitem[{{Weygand} {et~al.}(2011)}]{WeygandEA11}
{Weygand}, J. M., {Matthaeus}, W. H., {Dasso}, S., et al. 2011, 
\jgr, 116, A08102

\bibitem[{{Wijsen} {et~al.}(2019)}]{WijsenEA19}
Wijsen, N., Aran, A., Pomoell, J., et al. 2019, \aap, 622, A28

\bibitem[{{Wu} \& {Qin}(2018)}]{WuAQin18}
{Wu}, S.-S., \& {Qin}, G. 2018, \jgr: Space Physics, 123, 76

\bibitem[{{Yu et~al.}(2010)}]{YuEA10}
{Yu}, X. X., {Lu}, H., {Le}, G. M., et al. 2010, \solphys, 263, 223

\bibitem[{{Zank et~al.}(2018)}]{ZankEA18}
Zank, G. P., Adhikari, L., Zhao, L.-L., et al. 2018, \apj, 869, 23

\bibitem[{{Zank} \& {Matthaeus}(1992)}]{ZankEA92}
{Zank}, G. P., \& {Matthaeus}, W. H. 1992, \jgr, 97, 17189

\bibitem[{{Zank} \& {Matthaeus}(1993)}]{ZankEA93}
{Zank}, G. P., \& {Matthaeus}, W. H. 1993, PhFlA, 5, 257

\bibitem[{{Zhang} \& {Burlaga}(1988)}]{ZhangEA88}
{Zhang}, G., \& {Burlaga}, L. F. 1988, \jgr, 93, 2511

\bibitem[{{Zhang}(1999)}]{Zhang99}
{Zhang}, M. 1999, \apj, 513, 409

\bibitem[{{Zhang} {et~al.}(2009){Zhang}, {Qin}, \& {Rassoul}}]{ZhangEA09}
{Zhang}, M., {Qin}, G., \& {Rassoul}, H. 2009, 
\apj, 692, 109

\bibitem[{Zhao} {et~al.}(2018)]{ZhaoEA18}
Zhao, L.-L., Adhikari, L., Zank, G. P., et al. 2018, 
\apj, 856, 94

\bibitem[{{Zuo} {et~al.}(2011)}]{ZuoEA11}
{Zuo}, P., {Zhang}, M., {Gamayunov}, K., et al. 2011, 
\apj, 738, 168

\end{thebibliography}
\end{document}